\documentclass[prd,aps,showpacs,secnumarabic,superscriptaddress]{revtex4}
\usepackage[dvips]{graphicx,color}
\usepackage{bm}
\def\al{\alpha}

\def\ga{\gamma}
\def\de{\delta}

\def\ep{\varepsilon}

\def\ds{\displaystyle}

\def\slashed#1{#1\hspace{-5pt}/}
\def\Tr{{\rm Tr}}
\def\Dot{\!\cdot\!}
\begin{document}
\title{{ $\bm\ga\bm\ga$} and $\bm g\bm g$ decay rates for equal mass heavy quarkonia}
\author{James T. Laverty}
\email{laverty1@msu.edu}
\affiliation{Department of Physics and Astronomy, Michigan State University, East Lansing, Michigan 48824, USA} 
\author{Stanley F. Radford}
\email{sradford@brockport.edu}
\affiliation{Department of Physics, The College at Brockport, State University of New York, \\ Brockport, New York 14420, USA} 
\author{Wayne W. Repko}
\email{repko@pa.msu.edu}
\affiliation{Department of Physics and Astronomy, Michigan State University, East Lansing, Michigan 48824, USA} 
\date{\today}

\begin{abstract}
We present a calculation of the two-photon and two-gluon widths for the equal mass quarkonium states $^1S_0$, $^3P_0$ and $^3P_2$ of the charmonium and upsilon systems. The approach taken is based on using the full relativistic $q\bar{q}\to\ga\ga$ amplitude together with a wave function derived from the instantaneous Bethe-Salpeter equation. Momentum space radial wave functions obtained from an earlier fit of the charmonium and upsilon spectra are used to evaluate the necessary integrals. 
\end{abstract}
\maketitle

\section{Introduction}
Equal-mass quarkonia are eigenstates of the charge conjugation operator $C$ with eigenvalues $C=(-1)^{L+S}$. As such, the $^1S_0$, $^3P_0$ and $^3P_2$ levels of charmonium and the upsilon system can decay into two photons. These same states can also decay into two gluons, which accounts for a substantial portion of the hadronic decays for states below the $c\bar{c}$ or $b\bar{b}$ threshold. The two-photon decays of these states have been the subject of numerous studies aimed at further understanding the accuracy of theoretical models of the charmonium and upsilon systems based on the available data. Among the approaches used to study these decays are: the decomposition of the quark-antiquark annihilation amplitude into its two-component form \cite{GJR}, the use of a covariant light-front formalism \cite{HG}, the use of non-relativistic potential model techniques \cite{AM,LS}, the application of heavy-quark spin symmetry and effective Lagrangians \cite{LP1,LP2,LP3}, the use of a model with an AdS/QCD inspired potential \cite{G}, the application of relativistic two-body techniques \cite{HLTC,M,EFG,KLW}, the use of QCD sum rules \cite{NOSVVZ} and {\it ab inito} calculations using lattice QCD \cite{DE}. 

In the calculation of the two gamma widths that follows, we wish to include the relativistic and QCD effects in the wave functions that are used to compute the two photon decay amplitudes. We do this by making use of the variational wave functions obtained in \cite{RR}. These wave functions were computed using a semi-relativistic model containing both the $v^2/c^2$ and one-loop QCD corrections to the potential and optimized to provide an accurate description of the $c\bar{c}$ and $b\bar{b}$ quarkonium spectra. One of the parameters determined in this process is the renormalization scale appropriate to the particular quarkonium system. We use the resulting radial  functions to construct the individual $^1S_0$, $^3P_0$ and $^3P_2$ wave functions with the proper spin dependence obtained from a decomposition of the instantaneous Salpeter equation. One objective of taking this approach is to investigate how the inclusion of the full wave function information compares with the practice of using the square of the radial wave function $|R_{n0}(0)|^2$ for $s$-states or $|R'_{n1}(0)|^2$ for $p$-states. This much can be accomplished by using the expression for the invariant amplitude, ${\cal M}$, which can be derived using the instantaneous Salpeter wave function, $\phi(\vec{p})$  \cite{HLTC,M,EFG,KLW}, 
\begin{equation} \label{amplitude}
{\cal M}=e^2\int \!\!d^3p\,\Tr\,\left[C^{-1}\slashed{\ep}'^{*} S_F(p-k)  \slashed{\ep}^*\phi(\vec{p}) + C^{-1}\slashed{\ep}^* S_F(p-k') \slashed{\ep}'^{*}\phi(\vec{p})\right]\,.
\end{equation}
Here, $C$ is the charge conjugation matrix, $k$ and $k'$ are the photon momenta, $\ep$ and $\ep'$ are the photon polarization vectors, $S_F(p)$ is the quark propagator and $p_\mu=(\vec{p},i\sqrt{\vec{p}^{\,2}+m^2})=(\vec{p},iE)$, with $m$ denoting the quark mass. The wave function $\phi(\vec{p})$ is the Fourier transform of the instantaneous position space wave function $\psi(\vec{x})$ and its radial portion is obtained from the variational calculation in Ref.\cite{RR}.

There are also QCD corrections to the two gamma widths. Because of color conservation, these corrections are finite at the one-loop level. They could, in principle, be calculated by extending Eq.\,(\ref{amplitude}) as a perturbative series in the QCD coupling $\al_S$, but the one-loop corrections are known \cite{BCER,BCGR,KMRR,LC} and will be included in the final results.

In the next section, we construct the necessary wave functions, evaluate the trace and calculate the widths $\Gamma_{\ga\ga}$ for the $^1S_0$, $^3P_0$ and $^3P_2$ states of the $c\bar{c}$ and $b\bar{b}$ systems. We conclude with a numerical evaluation of the two-photon widths and some comments on the two-gluon widths $\Gamma_{\bm g\bm g}$.
\section{Evaluation of the Bound-State Decays}

The wave functions for the solution of the instantaneous Salpeter equation can be decomposed into spin singlet and spin triplet forms, which are \cite{Hostler:1980af}
\begin{equation}
^1\phi(\vec{p})=\frac{1}{2\sqrt{2}}\left[-\frac{i}{E}\vec{p}\Dot\vec{\ga}\ga_4+ \frac{m}{E}\ga_4 +1\right]\ga_5P(\vec{p})C\,,
\end{equation}
for the singlet states and 
\begin{eqnarray}
^3\phi(\vec{p})&=&\frac{1}{2\sqrt{2}}\left[-i\frac{1}{E}\vec{p}\Dot \vec{V}(\vec{p}) +\vec{\ga}\Dot\left(\vec{V}(\vec{p})-\frac{\vec{p}\,\vec{p}\Dot \vec{V}(\vec{p})}{E(E+m)}\right)\right. \nonumber \\ 
&&\left.+i\vec{\al}\Dot\left(\frac{m}{E}\vec{V}(\vec{p})+\frac{\vec{p}\,\vec{p}\Dot \vec{V}(\vec{p})}{E(E+m)}\right)+i\frac{1}{E}\vec{\ga}\ga_5\Dot\left(\vec{p}\times \vec{V}(\vec{p})\right)\right]C\,,
\end{eqnarray}
for the triplet states. These states have the conventional normalization
\begin{equation}
\int d^{\,3}p P^*(\vec{p})P(\vec{p})=1\qquad \int d^{\,3}p\, \vec{V}^*(\vec{p})\Dot \vec{V}(\vec{p})=1\,,
\end{equation}
where
\begin{equation}\label{wavefunctions}
\begin{array}{lcc} 
P(\vec{p})= \frac{\ds 1}{\sqrt{\ds 4\pi}}\,\phi_{n0}(p)  & {\rm for}  & ^1S_0 \\ [8pt]
\vec{V}(\vec{p})=\frac{\ds 1}{\sqrt{\ds 4\pi}}\hat{p}\,\phi_{n1}(p)   & {\rm for}  & ^3P_0 \\ [8pt]
V_i(\vec{p})=\sqrt{\frac{\ds 3}{\ds 4\pi}}\xi^{(M)}_{ij}\,\hat{p}_j\phi_{n1}(p) & {\rm for} & ^3P_2\,.
\end{array}
\end{equation}
The spin-two polarization vectors $\xi^{(M)}_{ij}$ satisfy $\xi^{(M)}_{ij}=\xi^{(M)}_{ji}$ and $\xi^{(M)}_{ii}=0$.

The explicit form of the $q\bar{q}\to\ga\ga$ amplitude in Eq.\,(\ref{amplitude}) is 
\begin{equation}
{\cal A}=e^2e_q^2\left[\frac{\slashed{\ep}'^*[m-i(\slashed{p}-\slashed{k})] \slashed{\ep}^*}{-2p\Dot k}+\frac{\slashed{\ep}^* [m-i(\slashed{p}-\slashed{k}')]\slashed{\ep}'^*}{-2p\Dot k'}\right]\,,
\end{equation}
where $e$ is the proton charge and $e_q$ the fractional quark charge.
With the aid of the wave functions in Eq.\,(\ref{wavefunctions}), the traces in Eq.\,(\ref{amplitude}) can be evaluated rather straightforwardly for the spin $0^\mp$ states. In the quarkonium rest frame, including the factors $1/2\omega$ from the photon normalization and $1/(2\pi)^{3/2}$ from the Fourier transform of the position space wave function, the results are, 
\begin{equation} \label{1S0}
{\cal M}(^1S_0)=e^2e_q^2\frac{1}{8\pi^2}\frac{m}{\omega}\int \!\!d^{\,3}p 
\frac{\hat{k}\Dot(\ep'^*\times\ep^*)} {[(\vec{p}\Dot\hat{k})^2-E^2]}\phi_{n0}(p)\,,
\end{equation}
for the $^1S_0$ state, and
\begin{equation} \label{3P0}
{\cal M}(^3P_0)=e^2e_q^2\frac{1}{8\pi^2}\frac{m}{\omega^2}\int \!\!d^{\,3}p \frac{1}{p\,E} \frac{[\,\omega(\vec{p}\Dot\hat{k})^2\ep'^*\Dot\ep^*+ 2E\vec{p}\Dot\ep'^*\vec{p}\Dot\ep^* \,]}{[(\vec{p}\Dot\hat{k})^2-E^2]} \phi_{n1}(p)\,,
\end{equation}
for the $^3P_0$ state. The evaluation of ${\cal M}(^3P_2)$ is somewhat more tedious, yielding
\begin{eqnarray}
{\cal  M}(^3P_2) &=& e^2e_q^2\frac{\sqrt{3}}{8\pi^2}\frac{1} {\omega^2} \int\!\!d^{\,3}p \frac{\xi^{(M)}_{ij}}{p\,[(\vec{p}\Dot\hat{k})^2-E^2]} \left[-E(\ep'^*_ip_j\ep^*\Dot\vec{p}+\ep'^*\Dot\vec{p}\,\ep^*_ip_j) -\omega\ep'^*\Dot\ep^*\,\vec{p}\Dot\hat{k}\,\hat{k}_ip_j\right. \nonumber \\
&& \left.\hspace{1.0in} +\frac{[\omega(\vec{p}\Dot\hat{k})^2\ep'^*\Dot\ep^*+2E\ep'^*\Dot\vec{p}\,\ep^*\Dot\vec{p}\,] \,p_ip_j}{E(E+m)}\right]\phi_{n1}(p)\,,
\end{eqnarray}
where we have dropped a term involving $p_ip_jp_k$ that vanishes upon solid angle integration.

To complete the calculation of the amplitude, we need to evaluate the angular integrals
\begin{eqnarray}
\int\!\!d\Omega_{\vec{p}}\frac{1}{[(\vec{p}\Dot\hat{k})^2-E^2]} &=& A_0 \,,\label{A0}\\
\int\!\!d\Omega_{\vec{p}}\frac{p_i\,p_j}{[(\vec{p}\Dot\hat{k})^2-E^2]} &=& A_1\hat{k}_i\hat{k}_j+A_2\de_{ij}\,,\label{A1A2} \\
\int\!\!d\Omega_{\vec{p}}\frac{p_i\,p_j\,p_k\,p_\ell}{[(\vec{p}\Dot\hat{k})^2-E^2]} &=&B_1\hat{k}_i\hat{k}_j\hat{k}_k\hat{k}_\ell+B_2(\de_{ij}\hat{k}_k\hat{k}_\ell +\de_{ik}\hat{k}_j\hat{k}_\ell+\de_{i\ell}\hat{k}_j\hat{k}_k+\de_{jk}\hat{k}_i\hat{k}_{\ell} +\de_{j\ell}\hat{k}_i\hat{k}_k+\de_{k\ell}\hat{k}_i\hat{k}_j) \nonumber \\
&&+B_3(\de_{ij}\de_{k\ell}+\de_{ik}\de_{j\ell}+\de_{jk}\de_{i\ell})\,,\label{B1B2B3}
\end{eqnarray}
noting that angular integrals of this type with an odd number of $p_i$'s in the numerator vanish. The values of the coefficients $A_0$ through $B_3$ are given in the Appendix. In terms of these coefficients, the $0^\mp$ amplitudes, including a factor of $\sqrt{3}$ for color, are 
\begin{equation}
{\cal M}(^1S_0)= \frac{e^2e_q^2\sqrt{3}}{8\pi^2}\frac{\hat{k}\Dot (\ep'^*\times\ep^*)}{\omega}\int_0^\infty\!\!dp\,p^2mA_0(p)\phi_{n0}(p)\equiv \sqrt{3}\al e_q^2\frac{\hat{k}\Dot (\ep'^*\times\ep^*)}{\omega}I_0\,,
\end{equation}
for the $0^-$ state and 
\begin{equation}
{\cal M}(^3P_0)= \frac{\sqrt{3}e^2e_q^2}{8\pi^2}\frac{\ep'^*\Dot\ep^*}{\omega^2} \int_0^\infty\!\!dp\frac{mp}{E}[\,\omega A_1+(\omega+2E)A_2]\phi_{n1}(p)\equiv \sqrt{3}\al e_q^2\frac{\ep'^*\Dot\ep^*}{\omega^2}J_0\,,
\end{equation}
for the $0^+$ state. As with the traces, the $^3P_2$ integration over $d\Omega_{\vec{p}}$ is more complicated since it involves four $p_i$'s in the numerator. The result is \cite{GJRerror}
\begin{eqnarray}
{\cal M}(^3P_2)&=&e^2e_q^2\frac{3}{8\pi^2}\frac{\xi^{(M)}_{ij}} {\omega^2}\int_0^\infty\!\!dp\,p\left[E\left(-A_2+\frac{2}{E(E+m)}B_3\right) (\ep'^*_i\ep^*_j+\ep'^*_j\ep^*_i)\right. \nonumber \\ [4pt]
&&+\left.\left(-\omega(A_1+A_2)+\frac{\omega(B_1+5B_2+2B_3)+2EB_2}{E(E+m)} \right)\ep'^*\Dot\ep^*\,\hat{k}_i\hat{k}_j\right]\phi_{n1}(p)\nonumber \\
&\equiv& 3\al e_q^2\frac{\xi^{(M)}_{ij}} {\omega^2} \left[(\ep'^*_i\ep^*_j+\ep'^*_j\ep^*_i)I_2+\ep'^*\Dot\ep^*\,\hat{k}_i\hat{k}_j\,J_2\right]\,.
\end{eqnarray}

The calculation of the widths can be performed using the formula for the decay of a particle with spin $J$ into two photons,
\begin{equation}\label{width}
\Gamma_{\ga\ga}=\frac{1}{16\pi^2}\omega^2\frac{1}{2J+1}\sum_{M=-J}^J  \sum_{\rm pol}\int\!\!d\Omega_{\hat{k}}|{\cal M}|^2.
\end{equation}
We sum over the photon polarizations using the Coulomb gauge
\begin{equation}
\sum_{\rm pol}\ep^*(\hat{k})\ep(\hat{k})=\de_{ij}-\hat{k}_i\hat{k}_j\,,
\end{equation}
and over the spin-two projections using
\begin{equation}
\sum_{M=-2}^2\xi^{(M)*}_{ij}\xi^{(M)}_{mn}=\frac{1}{2}(\de_{im}\de_{jn}+\de_{in}\de_{jm}) -\frac{1}{3}\de_{ij}\de_{mn}\,.
\end{equation}
The resulting two-gamma widths are
\begin{eqnarray}
\Gamma_{\ga\ga}(^1S_0) &=& \frac{3\al^2e_q^4}{2\pi}|I_0|^2\left(1+\frac{\al_S}{\pi}\left(\frac{\pi^2}{3}- \frac{20}{3}\right)\right)\,, \label{width1}\\
\Gamma_{\ga\ga}(^3P_0) &=& \frac{3\al^2e_q^4}{2\pi\omega^2}|J_0|^2\left(1+\frac{\al_S}{\pi}\left(\frac{\pi^2}{3}- \frac{28}{3}\right)\right)\,,\label{width2} \\
\Gamma_{\ga\ga}(^3P_2) &=& \frac{3\al^2e_q^4}{5\pi\omega^2}\left(6|I_2|^2+|I_2-J_2|^2\right)\left(1- \frac{16}{3}\frac{\al_S}{\pi}\right)\,, \label{width3}
\end{eqnarray}
where conservation of energy requires that the photon energy $\omega$ satisfies $\omega=M_{q\bar{q}}/2$. The last factors in Eqs.\,(\ref{width1})-(\ref{width3}) include the one-loop QCD correction. The two-loop QCD correction to the decay rate has been examined in \cite{CM} and appears to be large. We have not attempted to include this correction in our evaluation of the decay amplitudes. 

\section{Results and Conclusions}

The integrals that remain in Eqs.\,(\ref{width1})-(\ref{width3}) were evaluated using the perturbative wave functions from \cite{RR} with the aid of Mathematica. In Table \ref{charmwidth}, our results for the charmonium system are compared with those of other relativistic two-body approaches, Ref.\,\cite{HLTC,M,EFG,KLW}.
\begin{table}[h]
\caption{A comparison of our charmonium results with those of Ref.\,\cite{HLTC,M,EFG,KLW} is shown. The data denoted by $^*$ were taken from Ref.\cite{Amsler} and those denoted by $^\dagger$ were taken from Ref.\cite{Ecklund:2008hg}. The column labeled $\Gamma_{\rm NR+QCD}$ is the value of the width calculated using our values of $|R_{n0}(0)|^2$ or $|R'_{n1}(0)|^2$ along with the QCD correction.  \label{charmwidth}}
\vspace{2pt}
\begin{tabular}{ |r|r|r|r|r|r|r|r| }
\hline
\multicolumn{1}{|c|}{Decay} & \multicolumn{1}{c|}{Our} & \multicolumn{1}{c|}{Ref.\,\cite{HLTC}} & \multicolumn{1}{c|}{Ref.\,\cite{M}}& \multicolumn{1}{c|}{Ref.\,\cite{EFG}}& \multicolumn{1}{c|}{Ref.\,\cite{KLW}}& \multicolumn{1}{c|}{$\Gamma_{\rm NR+QCD}$} & \multicolumn{1}{c|}{$\Gamma_{\rm Exp}$} \\ \hline
$\eta_{c} \rightarrow \gamma \gamma$ & 5.09 keV & 5.5 keV & 3.5 keV & 5.5 keV & 7.14 keV & 13.1 keV & 7.2 $ \pm $0.9 keV$^*$\\
$\chi_{c0} \rightarrow \gamma \gamma $& 2.02 keV &  & 1.39 keV & 2.9 keV & & 5.35 keV         & 2.53 $ \pm $ 0.45 keV$^\dagger$\\
$\chi_{c2} \rightarrow \gamma \gamma$ & 0.46 keV &  & 0.44 keV & 0.50 keV & & 1.55 keV         & 0.60 $ \pm $ .08 keV$^\dagger$\\
$\eta^{'}_{c} \rightarrow \gamma \gamma$ & 2.63 keV & 2.1 keV & 1.38 keV & 1.8 keV & 4.44 keV & 10.5 keV & $<\,7.0\pm 3.5$ keV$^*$\\
\hline
\end{tabular}
\end{table}
The corresponding comparsion for the upsilon system system is given in Table \ref{upsilonwidth}.

\begin{table}[h]
\caption{A comparison of our upsilon results with those of Ref.\,\cite{HLTC,M,EFG,KLW} is shown. The column labeled $\Gamma_{\rm NR+QCD}$ is the value of the width calculated using $|R_{n0}(0)|^2$ or $|R'_{n1}(0)|^2$ along with the QCD correction. \label{upsilonwidth}}
\begin{tabular}{ |r|r|r|r|r|r|r| }
\hline
\multicolumn{1}{|c|}{Decay} & \multicolumn{1}{c|}{Our} & \multicolumn{1}{c|}{Ref.\,\cite{HLTC}} & \multicolumn{1}{c|}{Ref.\,\cite{M}}& \multicolumn{1}{c|}{Ref.\,\cite{EFG}}& \multicolumn{1}{c|}{Ref.\,\cite{KLW}}& \multicolumn{1}{c|}{$\Gamma_{\rm NR+QCD}$} \\
\hline
$\eta_{b} \rightarrow \gamma \gamma$ & 0.30 keV & 0.45 keV & 0.22 keV & 0.35 keV & 0.38 kev & 0.55 keV \\
$\chi_{b0} \rightarrow \gamma \gamma$ & 32.9 eV & & 24.0 eV & 38.0 eV & & 58.4 eV \\
$\chi_{b2} \rightarrow \gamma \gamma$ & 7.19 eV & & 5.6 eV & 8.0 eV & & 9.85 eV \\
$\eta^{'}_{b} \rightarrow \gamma \gamma$ & 0.14 keV & 0.21 keV & 0.11 keV & 0.15 keV & 0.19 keV & 0.20 keV \\
$\chi^{'}_{b0} \rightarrow \gamma \gamma$ & 34.1 eV & & 26.0 eV & 29.0 eV & & 68.3 eV\\
$\chi^{'}_{b2} \rightarrow \gamma \gamma$ & 7.59 eV & & 6.8 eV & 6.0 eV & & 11.5 eV \\
$\eta^{''}_{b} \rightarrow \gamma \gamma$ & 0.10 keV & & 0.084 keV & 0.10 keV & & 0.22 keV \\
\hline
\end{tabular}

\end{table}

For charmonium there is reasonable agreement between our results and the others of this type and they compare favorably with experiment. The differences between these results can be traced to the form of the static $q\bar{q}$ potential used in the solution of the two-body equation of motion. As a group the results are smaller than the non-relativistic widths including the one-loop QCD correction. The two-gamma decay rates for the upsilon system have not been measured and the $\eta_b$ ground state has only recently been observed \cite{Aubert}. The widths obtained using the non-relativistic results modified by QCD corrections are larger than our calculated widths or those found when using relativistic two-body formalisms. Note that, in this formalism, knowledge of the $^1S_0$ masses is not necessary for the calculation of their $\ga\ga$ widths since the only mass that occurs in Eq.\,(\ref{width1}) is the quark mass, which we obtain from Ref.\cite{RR}. This fact could account for the close agreement of our $\Gamma_{\ga\ga}(^1S_0)$ results with those of Ref.\cite{KLW}, even though these authors use a very different potential to get their radial wave functions. 

Finally, we can convert our results from two-photon decays to two-gluon decays by replacing $3\al^2\,e_q^4$ by $2\al_S^2/3$ in Eqs.\,(\ref{width1})\,-\,(\ref{width3}). This decay process accounts for a substantial portion of hadronic decays for states below $c\bar{c}$ or $b\bar{b}$ threshold. There are, however, significant radiative corrections as well as contributions from three-gluon decays and, thus, the two gluon mode does not tell the whole story. Our results for these two-gluon decays are given in Tables \ref{charmhadronic} and \ref{upsilonhadronic}. From Table
 III, it can be seen that the two-gluon widths are smaller than the hadronic widths of the charmonium states.
\begin{table}[h]
\caption{\label{charmhadronic}}
\begin{tabular}{ |r| r | r | }
\hline
\multicolumn{1}{|c|}{Decay} & \multicolumn{1}{c|}{$\Gamma_{\rm Th}$} & \multicolumn{1}{c|}{$\Gamma_{\rm Exp}$} \\ \hline
$\eta_{c} \rightarrow gg$ & 15.70 MeV &  $26.7\pm 3.0$ MeV \\
$\chi_{c0} \rightarrow gg$ & 4.68 MeV &  $10.2\pm 0.7$ MeV \\
$\chi_{c2} \rightarrow gg$ & 1.72 MeV &  $2.03\pm 0.12$ MeV  \\
$\eta^{'}_{c} \rightarrow gg$ & 8.10 MeV & $14\pm 7$ MeV \\
\hline
\end{tabular}

\end{table}

\begin{table}[h]
\caption{\label{upsilonhadronic}}
\begin{tabular}{ |r| r | }
\hline
\multicolumn{1}{|c|}{Decay} & \multicolumn{1}{c|}{$\Gamma_{\rm Th}$}  \\ 
\hline
$\eta_{b} \rightarrow gg$ & 11.49 MeV  \\
$\chi_{b0} \rightarrow gg$ & 0.96 MeV  \\
$\chi_{b2} \rightarrow gg$ & 0.33 MeV  \\
$\eta^{'}_{b} \rightarrow gg$ & 5.16 MeV  \\
$\chi^{'}_{b0} \rightarrow gg$ & 0.99 MeV \\
$\chi^{'}_{b2} \rightarrow gg$ & 0.35 MeV \\
$\eta^{''}_{b} \rightarrow gg$ & 3.80 MeV \\
\hline
\end{tabular}

\end{table}

\begin{acknowledgments}
This work was supported in part by the National Science Foundation under Grant PHY-0555544.
\end{acknowledgments}

\appendix
\section{Angular Integral Coefficients}
The coefficients $A_0$ through $B_3$ occurring in Eqs.\,(\ref{A0})\,-\,(\ref{B1B2B3}) are presented below \cite{GJR1}.
\begin{eqnarray}
A_0 &=& \frac{2\pi}{Ep}\ln \left(\frac{E-p}{E+p}\right)\,, \\
A_1 &=& \frac{\pi}{Ep}\left[(3E^2-p^2)\ln\left(\frac{E-p}{E+p}\right) + 6Ep\right]\,, \\
A_2 &=& -\frac{\pi}{Ep}\left[m^2\ln\left(\frac{E-p}{E+p}\right)+2Ep\right]\,, \\
B_1 &=& \frac{\pi}{12Ep} \left[(105E^4-90E^2p^2+9p^4)\ln\left(\frac{E-p}{E+p}\right)+(210E^3p-110Ep^3)\right]\,,\\
B_2 &=&-\frac{\pi}{12Ep} \left[(15E^4-18E^2p^2+3p^4)\ln\left(\frac{E-p}{E+p}\right)+(30E^3p-26Ep^3)\right]\,,\\
B_3  &=& \frac{\pi}{12Ep} \left[3m^4\ln\left(\frac{E-p}{E+p}\right)+(6E^3p-10Ep^3)\right]\,.
\end{eqnarray}


\begin{thebibliography}{99}
\bibitem{GJR} S. N. Gupta, J. M. Johnson, and W. W. Repko, Phys. Rev. D {\bf 54}, 2075 (1996).
\bibitem{HG} C.~W.~Hwang and R.~S.~Guo, arXiv:1005.2811 [hep-ph].
\bibitem{AM} M.~R.~Ahmady and R.~R.~Mendel, Phys.\ Rev.\  D {\bf 51}, 141 (1995) [arXiv:hep-ph/9401315].
\bibitem{LS} O.~Lakhina and E.~S.~Swanson, Phys.\ Rev.\  D {\bf 74}, 014012 (2006) [arXiv:hep-ph/0603164].
\bibitem{LP1} J.~P.~Lansberg and T.~N.~Pham, Phys. Rev. D {\bf 74}, 034001 (2006).
\bibitem{LP2} J.~P.~Lansberg and T.~N.~Pham, Phys.\ Rev.\  D {\bf 75}, 017501 (2007) [arXiv:hep-ph/0609268].
\bibitem{LP3} J.~P.~Lansberg and T.~N.~Pham, Phys.\ Rev.\  D {\bf 79}, 094016 (2009) [arXiv:0903.1562 [hep-ph]].
\bibitem{G} F.~Giannuzzi, Phys.\ Rev.\  D {\bf 78}, 117501 (2008) [arXiv:0810.2736 [hep-ph]].
\bibitem{HLTC} K.~T.~Chao, H.~W.~Huang, J.~H.~Liu and J.~Tang, Phys.\ Rev.\  D {\bf 56}, 368 (1997) [arXiv:hep-ph/9601381].
\bibitem{M} C.~R.~Munz, Nucl.\ Phys.\  A {\bf 609}, 364 (1996) [arXiv:hep-ph/9601206].
\bibitem{EFG} D.~Ebert, R.~N.~Faustov and V.~O.~Galkin, Mod.\ Phys.\ Lett.\  A {\bf 18}, 601 (2003) [arXiv:hep-ph/0302044].
\bibitem{KLW} C.~S.~Kim, T.~Lee and G.~L.~Wang, Phys.\ Lett.\  B {\bf 606}, 323 (2005) [arXiv:hep-ph/0411075]
\bibitem{NOSVVZ} V.~A.~Novikov, L.~B.~Okun, M.~A.~Shifman, A.~I.~Vainshtein, M.~B.~Voloshin and V.~I.~Zakharov, Phys.\ Rept.\  {\bf 41}, 1 (1978).
\bibitem{DE} J.~J.~Dudek and R.~G.~Edwards, Phys. Rev. Lett. {\bf 97}, 172001 (2006).
\bibitem{RR} S.~F.~Radford and W.~W.~Repko, Phys. Rev. D {\bf 75}, 074031 (2007) [arXiv:hep-ph/0701117].
\bibitem{BCER} R.~Barbieri, G.~Curci, E.~d'Emilio and E.~Remiddi, Nucl. Phys. {\bf B154}, 535 (1979).
\bibitem{BCGR} R.~Barbieri, M.~Caffo, R.~Gatto and E.~Remiddi, Phys. Lett. {\bf 95B}, 93 (1980); Nucl. Phys. {\bf B192}, 61 (1981).
\bibitem{KMRR} W.~Kwong, P.~B.~Mackenzie, R. Rosenfeld and J.~L.~Rosner, Phys. Rev. D {\bf 37}, 3210 (1988).
\bibitem{LC} B.-Q.~Li and K.-T.~Chao, [arXiv:0903.5506].
\bibitem{Hostler:1980af} L.~C.~Hostler and W.~W.~Repko, Annals Phys.\  {\bf 130}, 329 (1980).
\bibitem{CM} A.~Czarnecki and K. Melnikov, Phys. Lett. {\bf B519}, 212 (2001).
\bibitem{Amsler} C.~Amsler, {\it et al.}, Phys. Lett. {\bf B667}, 1 (2008).
\bibitem{Ecklund:2008hg} K.~M.~Ecklund {\it et al.}  [CLEO Collaboration], arXiv:0803.2869 [hep-ex].
\bibitem{Aubert} C.~Aubert, {\it et al.}, arXiv:0807.1086.
\bibitem{GJRerror} The definitions of $I_2$ and $J_2$ contain an additional factor of $E$ in the numerator missing in the corresponding definitions of $I_3$ and $I_4$ in Ref.\cite{GJR}. This is merely a typo - the integrals were evaluated with the correct expressions.
\bibitem{GJR1} These results were independently checked by us, but can be obtained from those in Ref.\cite{GJR}.
\end{thebibliography}
\end{document}